\documentstyle[12pt,aps]{revtex}
\begin{document}
\draft
\author{Yishi Duan, Libin Fu\thanks{%
Corresponding author. E-mail: itp2@lzu.edu.cn} and Guang Jia}
\address{Institute of Theoretical Physics,Lanzhou University,Lanzhou, Gansu,\\
730000, P.R. China}
\title{Topological tensor current of $\tilde p$-branes in the $\phi $-mapping theory}
\date{\today }
\maketitle

\begin{abstract}
\begin{center}
{\bf Abstract}
\end{center}

We present a new topological tensor current of $\tilde p$-branes by making
use of the $\phi $-mapping theory. It is shown that the current is
identically conserved and behave as $\delta (\vec \phi ),$ and every
isolated zero of the vector field $\vec \phi (x)$ corresponds to a
`magnetic' $\tilde p$-brane. Using this topological current, the generalized
Nambu action for multi $\tilde p$-branes is given, and the field strength $F$
corresponding to this topological tensor current is obtained. It is also
shown that the `magnetic' charges carried by $\tilde p$-branes are
topologically quantized and labeled by Hopf index and Brouwer degree, the
winding number of the $\phi $-mapping.
\end{abstract}

\section{Introduction}

Extended objects with $p$ spatial dimensional, known as `branes', play an
essential role in revealing the non-perturbative structure of the
superstring theories and $M$-theories \cite{b1,b2,b3,b4}. Antisymmetric
tensor gauge fields have been widely studied in the theories of $p$-branes%
\cite{a1,a2,a3}. In the context of the effective $D=10$ or $D=11$
supergravity theory a $p$-brane is a $p$-dimensional extended source for a $%
(p+2)$-form gauge field strength $F.$ It is well-known that the $(p+2)$-form
strength $F$ satisfies the field equation%
$$
\nabla _\mu F^{\mu \mu _1\cdots \mu _{p+1}}=j^{\mu _1\cdots \mu _{p+1}} 
$$
where $j^{\mu _1\cdots \mu _{p+1}}$is a $(p+1)$-form tensor current and
corresponding to electric source, and the dual field strength $^{*}F$
satisfies%
$$
\nabla _\mu ~^{*}F^{\mu \mu _1\cdots \mu _{\tilde p+1}}=\tilde j^{\mu
_1\cdots \mu _{\tilde p+1}} 
$$
in which $\tilde j^{\mu _1\cdots \mu _{\tilde p+1}}$ is a $(\tilde p+1)$%
-form tensor current and corresponding to magnetic source \cite{strom}\cite
{hull}\cite{duff}.

The $\phi $-mapping theory proposed by Prof. Duan\cite{DuanGe,DuanLiu} is
important in studying the topological invariant and topological structure of
physics systems and has been used to study topological current of magnetic
monopole \cite{DuanGe}, topological string theory \cite{DuanLiu},
topological structure of Gauss-Bonnet-Chern theorem \cite{DuanMeng2},
topological structure of the SU(2) Chern density\cite{fu} and topological
structure of the London equation in superconductor \cite{DuanZhangLi}. We
must pointed out that the $\phi $-mapping theory is also a powerful tools to
investigate the topological defects theory \cite{DuanZhangFeng,DuanZhang,DZF}%
, and here the vector field $\vec \phi $ is looked upon as the order
parameters of the defects.

In this paper, we present a new topological tensor current of `magnetic' $%
\tilde p$-branes by making use of the $\phi $-mapping theory. One shows that
the each isolated zero of the $d$-dimensional vector field $\vec \phi (x)$
corresponds to a $\tilde p$-brane$(\tilde p=D-d-1),$ and this current is
proved to be the general current density of multi $\tilde p$-branes. Using
this current, the generalized Nambu action for multi $\tilde p$-branes is
obtained. This topological tensor current will give rise to the inner
structure of the field strength $F$ including the contribution of the
`magnetic' $\tilde p$-branes. Finally, we show that the charges carried by
multi $\tilde p$-branes are topologically quantized and labeled by the Hopf
index and Brouwer degree, the winding number of the $\phi $-mapping.

\section{The topological tensor current of $\tilde p$-branes}

Let $X$ be a $D$-dimensional smooth manifold with metric tensor $g_{\mu \nu
} $ and local coordinates $x^\mu (\mu ,\nu =0,\cdots ,D-1)$ with $x^0=t$ as
time, and let $R^d$ be an Euclidean space of dimension $d<D.$ We consider a
smooth map $\phi :X\rightarrow R^d$ , which gives a $d$-dimensional smooth
vector field on $X$%
\begin{equation}
\phi ^a=\phi ^a(x),\quad a=1,2,\cdots ,d. 
\end{equation}
The direction unit field of $\vec \phi (x)$ can be expressed as 
\begin{equation}
\label{unit}n^a=\frac{\phi ^a}{||\phi ||},\quad \quad ||\phi ||=\sqrt{\phi
^a\phi ^a}. 
\end{equation}
In the $\phi $-mapping theory, to extend the theory of magnetic monopoles%
\cite{DuanGe} and the topological string theory\cite{DuanLiu}, we present a
new topological tensor current, with the unit `magnetic' charge $g_m$,
defined as 
$$
\tilde j^{\mu _1\cdots \mu _{D-d}}=\frac{g_m}{A(S^{d-1})(d-1)!}(\frac 1{%
\sqrt{g}})\in ^{\mu _1\cdots \mu _{D-d}\mu _{D-d+1}\mu _{D-d+2}\cdots \mu
_D} 
$$
\begin{equation}
\label{current1}\in _{a_1a_2\cdots a_d}\partial _{\mu
_{(D-d+1)}}n^{a_1}\partial _{\mu _{(D-d-2)}}n^{a_2}\cdots \partial _{\mu
_D}n^{a_d} 
\end{equation}
where $g$ is the determinant of the metric tensor $g_{\mu \nu }$. Obviously,
this `magnetic' tensor current is identically conserved, 
\begin{equation}
\nabla _{\mu _1}\tilde j^{\mu _1\cdots \mu _{D-d}}=0,\quad i=1,\cdots ,D-d. 
\end{equation}
From (\ref{unit}) we have 
\begin{equation}
\label{ded}\partial _\mu n^a=\frac 1{||\phi ||}\partial _\mu \phi ^a+\phi
^a\partial _\mu (\frac 1{||\phi ||}) 
\end{equation}
\begin{equation}
\label{ded1}\frac \partial {\partial \phi ^a}(\frac 1{||\phi ||})=-\frac{%
\phi ^a}{||\phi ||^3} 
\end{equation}
Using the above expressions, the general tensor current can be rewritten as 
$$
\tilde j^{\mu _1\cdots \mu _{D-d}}=\frac{g_m}{A(S^{d-1})(d-1)!(d-2)}(\frac 1{%
\sqrt{g}})\in ^{\mu _1\cdots \mu _{D-d}\mu _{D-d+1}\cdots \mu _{D\ }}\in _{\
a_1\cdots a_d} 
$$
\begin{equation}
\partial _{\mu _{(D-d+1)}}\phi ^a\partial _{\mu _{(D-d-2)}}\phi ^{a_2}\cdots
\partial _{\mu _D}\phi ^{a_d}\frac \partial {\partial \phi ^a}\frac \partial
{\partial \phi ^{a_1}}(\frac 1{||\phi ||^{d-2\ }}). 
\end{equation}
If we define a generalized Jacobians tensor as 
\begin{equation}
\label{jaco}\in ^{a_1\cdots a_d}J^{\mu _1\cdots \mu _{D-d}}(\frac \phi x%
)=\in ^{\mu _1\cdots \mu _{D-d}\mu _{D-d+1}\mu _{D-d+2}\cdots \mu
_D}\partial _{\mu _{(D-d+1)}}\phi ^{a_1}\partial _{\mu _{(D-d-2)}}\phi
^{a_2}\cdots \partial _{\mu _D}\phi ^{a_d} 
\end{equation}
and make use of the generalized Laplacian Green function relation in $\phi $%
-space 
\begin{equation}
\label{delt}\frac \partial {\partial \phi ^a}\frac \partial {\partial \phi ^a%
}(\frac 1{||\phi ||^{d-2\ }})=\frac{4\pi ^{\frac d2}}{\Gamma (d/2-1)}\delta (%
\vec \phi ), 
\end{equation}
we obtain a $\delta $-function like tensor current\cite{DuanLiu} 
\begin{equation}
\label{current}\tilde j^{\mu _1\cdots \mu _{D-d}}=g_m\delta (\vec \phi
)J^{\mu _1\cdots \mu _{D-d}}(\frac \phi x)(\frac 1{\sqrt{g}}). 
\end{equation}
We find that $\tilde j^{\mu _1\cdots \mu _{D-d}}\neq 0$ only when $\phi =0$.
So, it is essential to discuss the solutions of the equations 
\begin{equation}
\phi ^a(x)=0,\quad a=1,\cdots ,d 
\end{equation}

Suppose that the vector field $\vec \phi (x)$ possesses $l$ isolated zeroes,
according to the deduction of Ref. \cite{DuanLiu} and the implicit function
theorem\cite{ggg}\cite{yang76}, when the zeroes are regular points of $\phi $%
-mapping, i.e. the rank of the Jacobian matrix $[\partial _\mu \phi ^a]$ is $%
d$, the solution of $\vec \phi (x)=0$ can be parameterized by 
\begin{equation}
\label{solut}x^\mu =z_i^\mu (u^1,u^2,\cdots ,u^{D-d}),\quad i=1,\cdots ,l, 
\end{equation}
where the subscript $i$ represents the $i$-th solution and the parameters $%
u=u(u^1,\cdots ,u^{D-d})$ span a $(D-d)$-dimensional submanifold of $X$,
denoted by $N_i$, which corresponds to a $\tilde p$-brane$(\tilde p=D-d-1)$
with spatial $\tilde p$-dimension and $N_i$ is its worldvolume. One see that
the tensor current $\tilde j^{\mu _1\cdots \mu _{D-d}}$ is not vanished only
on the worldvolume manifolds $N_i$ $(i=1,\cdots ,l),$ each of which
corresponds to a $\tilde p$-brane. Therefore, every isolated zero of $\vec 
\phi (x)$ on $X$ corresponds to a magnetic $\tilde p$-branes. These
`magnetic' $\tilde p$-branes had been formally discussed and not studied
based on the topology theory \cite{nepo,teit}. Here, we must pointed out
that the $\tilde p$-branes, sometimes, may be considered as topological
defects \cite{duff,tp}, in this case for our theory the vector field $\phi
^a(x)$ $(a=1,\cdots ,d)$ may be looked upon as the generalized order
parameters\cite{DZF} for $\tilde p$-branes.

In the following, we will discuss the inner structure of the topological
tensor current $\tilde j^{\mu _1\cdots \mu _{D-d}}$. It can be proved that
there exists a $d$-dimensional submanifold $M$ in $X$ with the parametric
equation 
\begin{equation}
\label{trcol}x^\mu =x^\mu (v^1,\cdots ,v^d),\quad \mu =1,\cdots ,D, 
\end{equation}
which is transversal to every $N_i$ at the point $p_i$ with 
\begin{equation}
\label{normal}g_{\mu \nu }\frac{\partial x^\mu }{\partial u^I}\frac{\partial
x^v}{\partial v^A}|_{p_i}=0,\quad I=1,\cdots ,D-d,\quad A=1,\cdots ,d. 
\end{equation}
This is to say that the equations $\vec \phi (x)=0$ have the isolated zero
points on $M$.

As we have pointed in Ref. \cite{DuanMeng2,fu}, the unit vector field
defined in (\ref{unit}) gives a Gauss map $n:\partial M_i\rightarrow S^{d-1}$%
, and the generalized Winding Number can be given by this Gauss map 
$$
W_i=\frac 1{A(S^{d-1})(d-1)!}\int_{\partial M_i}n^{*}(\in _{a_1\cdots
a_d}n^{a_1}dn^{a_2}\wedge \cdots \wedge dn^{a_d}) 
$$
$$
\qquad \qquad \qquad \quad =\frac 1{A(S^{d-1})(d-1)!}\int_{\partial M_i}\in
_{a_1\cdots a_d}n^{a_1}\partial _{A_2}n^{a_2}\cdots \partial
_{A_d}n^{a_d}dv^{A_2}\wedge \cdots \wedge dv^{A_d} 
$$
\begin{equation}
\qquad \qquad \quad \qquad =\frac 1{A(S^{d-1})(d-1)!}\int_{M_i}\in
^{A_1\cdots A_d}\in _{a_1\cdots a_d}\partial _{A_1}n^{a_1}\partial
_{A_2}n^{a_2}\cdots \partial _{A_d}n^{a_d}d^dv.
\end{equation}
where $\partial M_i$ is the boundary of the neighborhood $M_i$ of $p_i$ on $M
$ with $p_i\notin \partial M_i,$ $M_i\cap M_j=\emptyset .$ Then, by
duplicating the derivation of (\ref{current1}) from (\ref{current}), we
obtain 
\begin{equation}
\label{wind}W_i=\int_{M_i}\delta (\vec \phi (v))J(\frac \phi v)d^dv
\end{equation}
where $J(\frac \phi v)$ is the usual Jacobian determinant of $\vec \phi $
with respect to $v$%
\begin{equation}
\in ^{a_1\cdots a_d}J(\frac \phi v)=\in ^{A_1\cdots A_d}\partial
_{A_1}n^{a_1}\partial _{A_2}n^{a_2}\cdots \partial _{A_d}n^{a_d}.
\end{equation}
According to the $\delta $-function theory \cite{yangd77} and the $\phi $%
-mapping theory, we know that $\delta (\vec \phi (v))$ can be expanded as 
\begin{equation}
\label{ff}\delta (\vec \phi (v))=\sum_{i=1}^l\beta _i\eta _i\delta ^d(\vec v-%
\vec v(p_i))
\end{equation}
on $M,$ where the positive integer $\beta _i=|W_i|$ is called the Hopf index
of the map $v\rightarrow \vec \phi (v)$ and $\eta _i=sgn(J(\frac \phi v%
))|_{p_i}=\pm 1$ is the Brouwer degree\cite{DuanMeng2,DuanZhangLi}. One can
find the relation between the Hopf index $\beta _i,$ the Brouwer degree $%
\eta _i$, and the winding number $W_i$ 
\begin{equation}
W_i=\beta _i\eta _i,
\end{equation}
One see that the Eq. (\ref{ff}) is only the expansion of $\delta (\vec \phi
(x))$ on $M.$ In order to investigate the expansion of $\delta (\vec \phi
(x))$ on the whole manifold $X$, we must expand the $d$-dimensional $\delta $%
-function of the singular point in terms of the $\delta $-function on the
singular submanifold $N_i$ which had been given in Ref. \cite{yangd77}%
$$
\delta (N_i)=\int_{N_i}\delta ^D(x-z_i(u))\sqrt{g_u}d^{(D-d)}u,\quad
i=1,\cdots ,l 
$$
in which 
\begin{equation}
g_u=\det (g_{\mu \nu }\frac{\partial x^\mu }{\partial u^I}\frac{\partial
x^\nu }{\partial u^J}),\quad I,J=1,\cdots ,(D-d).
\end{equation}
Then, from Eqs. (\ref{ff}), and by considering the property of the $\delta $%
-function, one will obtain 
\begin{equation}
\label{redelt}\delta (\vec \phi (x))=\sum_{i=1}^l\beta _i\eta
_i\int_{N_i}\delta ^D(x-z_i(u))\sqrt{g_u}d^{(D-d)}u.
\end{equation}
Therefore, the general topological current of the $\tilde p$-branes can be
expressed directly as 
\begin{equation}
\label{recur}\tilde j^{\mu _1\cdots \mu _{D-d}}=(\frac 1{\sqrt{g}})J^{\mu
_1\cdots \mu _{D-d}}(\frac \phi x)\sum_{i=1}^l\beta _i\eta
_i\int_{N_i}\delta ^D(x-z_i(u))\sqrt{g_u}d^{(D-d)}u,
\end{equation}
which is a new topological current theory of $\tilde p$-branes based on the $%
\phi $-mapping theory.

If we define a Lagrangian as 
\begin{equation}
L=\sqrt{\frac 1{(D-d)!}g_{\mu _1\nu _1}\cdots g_{\mu _{(D-d)}\nu _{(D-d)}}%
\tilde j^{\mu _1\cdots \mu _{D-d}}\tilde j^{\nu _1\cdots \nu _{D-d}}}, 
\end{equation}
which is just the generalization of Nielsen's Lagrangian\cite{niel}, from
the above deductions, we can prove that 
\begin{equation}
L=(\frac 1{\sqrt{g}})\delta (\vec \phi (x)). 
\end{equation}
Then, the action takes the form 
\begin{equation}
\label{aa}S=\int_XL\sqrt{g}d^Dx=\int_X\delta (\vec \phi (x))d^Dx. 
\end{equation}
By substituting the formula (\ref{redelt}) into (\ref{aa}), we obtain an
important result%
$$
S=\int_X\sum_{i=1}^l\beta _i\eta _i\int_{N_i}\delta ^D(x-z_i(u))\sqrt{g_u}%
d^{(D-d)}ud^Dx 
$$
\begin{equation}
=\sum_{i=1}^l\beta _i\eta _i\int_{N_i}\sqrt{g_u}d^{(D-d)}u, 
\end{equation}
i.e. 
\begin{equation}
S=\sum_{i=1}^l\eta _iS_i, 
\end{equation}
where $S_i=\beta _i\int_{N_i}\sqrt{g_u}d^{(D-d)}u.$ This is just the
generalized Nambu action for multi $\tilde p$-branes$(\tilde p=D-d-1),$
which is the straightforward generalization of Nambu action for the string
world-sheet action\cite{nambu}. Here this action for multi $\tilde p$-branes
is obtained directly by $\phi $-mapping theory, and it is easy to see that
this action is just Nambu action for multi-strings when $D-d=2$ \cite
{DuanLiu}.

\section{The gauge field corresponding to the topological current}

In this section, we will study the antisymmetric tensor gauge field
corresponding to the topological tensor current presented in above section.
We know that $p$-branes naturally acts as the `electric' source of a rank $%
p+2$ field strength 
\begin{equation}
\label{st}F=dA, 
\end{equation}
where $A$ is a ($p+1)$-form as the tensor gauge potential and satisfies the
gauge transformation 
$$
A\rightarrow A+d\Lambda _p. 
$$
From Eq. (\ref{st}), one have the Bianchi identity 
\begin{equation}
\label{bi}dF\equiv 0. 
\end{equation}
And the `electric' current density associated with the source can be
expressed as 
\begin{equation}
\label{equat1}j^{\mu _1\cdots \mu _{p+1}}=\nabla _\mu F^{\mu \mu _1\cdots
\mu _{p+1}}. 
\end{equation}

Just as the usual Maxwell's equation, we know that Eqs. (\ref{st}), (\ref{bi}%
) and (\ref{equat1}) imply the presence of an `electric' charge, i.e. $p$%
-branes, but no `magnetic' source \cite{duff}.

Now, let us discuss the case when there exists the `magnetic' source. For
this case, one must introduce another $(p+2)$-form $G$ for the magnetic
source, and the field strength $F$ must be modified to 
\begin{equation}
\label{stt}F=dA+G, 
\end{equation}
which is the generalized field strength including the contribution of the
`magnetic' source, i.e. `magnetic' branes: $\tilde p$-branes with $\tilde p%
=D-p-4$.

To obtain the explicit expression for $G,$ let us consider the current
density corresponds to magnetic source which is given by 
\begin{equation}
\label{equat11}\tilde j^{\mu _1\cdots \mu _{\tilde p+1}}=\nabla _\mu
~^{*}F^{\mu \mu _1\cdots \mu _{\tilde p+1}}.
\end{equation}
Using (\ref{stt}) and (\ref{equat11}), we obtain 
\begin{equation}
\label{resultg}\tilde j^{\mu _1\cdots \mu _{(\tilde p+1)}}=\frac 1{\sqrt{g}}%
\partial _\mu (\sqrt{g}\frac{\in ^{\mu \mu _1\cdots \mu _{\tilde p+1}\mu _{%
\tilde p+2}\cdots \mu _{D-1}}}{\sqrt{g}}G_{\mu _{\tilde p+2}\cdots \mu
_{D-1}}),
\end{equation}
It has been pointed out in the above section that the current density of the
`magnetic' branes is a topological current given by Eq. (\ref{current1}),
which can be rewritten as 
$$
\tilde j^{\mu _1\cdots \mu _{D-d}}=\frac{g_m}{A(S^{d-1})(d-1)!}(\frac 1{%
\sqrt{g}})\partial _{\mu _{(D-d+1)}}(\in ^{\mu _1\cdots \mu _{D-d}\mu
_{D-d+1}\mu _{D-d+2}\cdots \mu _D} 
$$
\begin{equation}
\label{currentg}\in _{a_1a_2\cdots a_d}n^{a_1}\partial _{\mu
_{(D-d-2)}}n^{a_2}\cdots \partial _{\mu _D}n^{a_d})
\end{equation}
where $(D-d)=\tilde p+1,$ i.e. $\tilde p=D-d-1$. Comparing the Eq. (\ref
{resultg}) to (\ref{currentg}), we can obtain 
\begin{equation}
\label{gg}G_{\mu _1\cdots \mu _{d-1}}=\frac{(-1)^{(D-d)}g_m}{A(S^{d-1})(d-1)!%
}\in _{a_1a_2\cdots a_d}n^{a_1}\partial _{\mu _1}n^{a_2}\cdots \partial
_{\mu _{d-1}}n^{a_d},
\end{equation}
and 
\begin{equation}
\label{aaz}G=\frac{(-1)^{(D-d)}g_m}{A(S^{d-1})(d-1)!}\in _{a_1a_2\cdots
a_d}n^{a_1}dn^{a_2}\wedge \cdots \wedge dn^{a_d}.
\end{equation}

Of equal interest is the `magnetic' charge carried by the multi $\tilde p$%
-branes, which is given by 
\begin{equation}
\label{gg1}Q^M=\int_\Sigma \tilde j^{\mu _1\cdots \mu _{\tilde p+1}}\sqrt{g}%
d\sigma _{\mu _1\cdots \mu _{\tilde p+1}}
\end{equation}
where $\Sigma $ is a $d$-dimension$(d=p+3)$ hypersurface in $X,$ while $%
d\sigma _{\mu _1\cdots \mu _{\tilde p+1}}$is the convariant surface element
of $\Sigma $ \cite{y96}. From (\ref{equat11}) and (\ref{gg1}), it is easy to
prove that%
$$
Q^M=\int_{\partial \Sigma }F, 
$$
where $\partial \Sigma $ is the boundary of $\Sigma $ and a $(p+2)$%
-dimension hypersurface. Substituting (\ref{recur}) into (\ref{gg1}), we
have 
\begin{equation}
\label{gg2}Q^M=g_m\int_\Sigma J^{\mu _1\cdots \mu _{\tilde p+1}}(\frac \phi x%
)\sum_{i=1}^l\beta _i\eta _i\int_{N_i}\delta ^D(x-z_i(u))\sqrt{g_u}%
d^{(D-d)}ud\sigma _{\mu _1\cdots \mu _{\tilde p+1}},
\end{equation}
from (\ref{jaco}), and the relation 
$$
\frac 1{(\tilde p+1)!}\in ^{\mu _1\cdots \mu _{\tilde p+1}\nu _1\cdots \nu
_d}d\sigma _{\mu _1\cdots \mu _{\tilde p+1}}=dx^{\nu _1}\wedge \cdots \wedge
dx^{\nu _d}, 
$$
the expression (\ref{gg2}) can be rewritten as 
\begin{equation}
\label{gg3}Q^M=\dot g_0\int_{\phi (\Sigma )}\sum_{i=1}^l\beta _i\eta
_i\int_{N_i}\frac 1{\sqrt{g}}\delta ^D(x-z_i(u))\sqrt{g_u}%
d^{(D-d)}ud^{(d)}\phi .
\end{equation}
Since on the singular submanifold $N_i$ we have 
\begin{equation}
\phi ^a(x)|_{N_i}=\phi ^a(z_i^1(u),\cdots ,z_i^D(u))\equiv 0,
\end{equation}
which leads to 
\begin{equation}
\partial \phi ^a\frac{\partial x^\mu }{\partial u^I}|_{N_i}=0.
\end{equation}
Using this expression, one can prove 
\begin{equation}
J^{\mu _1\cdots \mu _{D-d}}(\frac \phi x)|_{\vec \phi =0}=\frac{\sqrt{g}}{%
\sqrt{g_u}}\in ^{I_1\cdots I_{(D-d)}}\frac{\partial x^{\mu _1}}{\partial
u^{I_1}}\cdots \frac{\partial x^{\mu _{(D-d)}}}{\partial u^{I_{(D-d)}}}.
\end{equation}
Then we obtain an useful formula 
\begin{equation}
\label{import}d^{(d)}\phi \sqrt{g_u}d^{(D-d)}u=\sqrt{g}d^Dx.
\end{equation}
By making use of the above formula and (\ref{gg3}), we finally get 
\begin{equation}
Q^M=g_m\sum_{i=1}^l\beta _i\eta _i\int_X\delta
^D(x-z_i(u))d^Dx=g_m\sum_{i=1}^l\beta _i\eta _i.
\end{equation}
The above expression shows that the $i$-th brane carries the `magnetic'
charge $Q_i^M=g_m\beta _i\eta _i=g_mW_i,$ which is topologically quantized
and characterized by Hopf index $\beta _i$ and Brouwer degree $\eta _i,$ the
winding number $W_i$ of the $\phi $-mapping.

\section{Conclusion}

In this paper the $\phi $-mapping theory is introduced to study the $\tilde p
$-branes theory, which is development of our former theories of magnetic
monopoles and topological strings. We present a new topological tensor
current of magnetic multi $\tilde p$-branes and discuss the inner structure
of this current in detail. It is shown that every isolated zero of the
vector field $\vec \phi $ (i.e. order parameters) is just corresponding to a
magnetic brane, $\tilde p$-brane$(\tilde p=D-d-1)$. The generalized Nambu
action for multi $\tilde p$-branes can be obtained directly in terms of this
topological current. The topological structure of the charges carried by $%
\tilde p$-branes shows that the magnetic charges is topologically quantized
and labeled by the Hopf index and Brouwer degree, the winding number of the $%
\phi $-mapping. The theory formulated in this paper is a new concept for
topological $\tilde p$-branes based on the $\phi $-mapping theory.

\section{Acknowledgment}

This work was supported by the National Natural Science Foundation of China
and Doctoral Science Foundation of China.

\end{document}